\documentclass[10pt,superscriptaddress,twocolumn,showpacs,showkeys]{revtex4}
\usepackage{amsmath}
\usepackage[pdftex]{graphicx}
\usepackage{subfigure}

\begin{document}

\title{On the stability of the Einstein Static Universe in Massive Gravity}
\author{Luca Parisi}
\email{parisi@sa.infn.it}
\affiliation{Dipartimento di Fisica "E.R.Caianiello", Universit\`a di Salerno, I-84084 Fisciano (Sa), Italy \\INFN, Sezione di Napoli, GC di Salerno, I-84084 Fisciano (Sa), Italy}
\author{Ninfa Radicella}
\email{ninfa.radicella@uab.cat}
\affiliation{Dipartimento di Fisica "E.R.Caianiello", Universit\`a di Salerno, I-84084 Fisciano (Sa), Italy \\INFN, Sezione di Napoli, GC di Salerno, I-84084 Fisciano (Sa), Italy}
\author{Gaetano Vilasi}
\email{vilasi@sa.infn.it}
\address{Dipartimento di Fisica "E.R.Caianiello", Universit\`a di Salerno, I-84084 Fisciano (Sa), Italy \\INFN, Sezione di Napoli, GC di Salerno, I-84084 Fisciano (Sa), Italy}

\begin{abstract}
We consider static cosmological solutions along with their stability properties in the framework of a recently proposed theory of massive gravity. We show that the modification introduced in the cosmological equations leads to several new solutions, only sourced by a perfect fluid, generalizing the Einstein Static Universe found in General Relativity. Using dynamical system techniques and numerical analysis, we show that the found solutions can be either neutrally stable or unstable against spatially homogeneous and isotropic perturbations. 

\end{abstract}

\maketitle
\section{Introduction}\label{intro}

The exact solution of Einstein's equations known as the Einstein Static (ES) Universe is a static closed Friedmann-Robertson-Walker model sourced by a perfect fluid and a cosmological constant (see \cite{Hawking:1973uf}). Its stability properties have been widely investigated. The ES Universe is unstable to homogeneous perturbations \cite{Edd:1930}, moreover it is always neutrally stable against small inhomogeneous vector and tensor perturbations and neutrally stable against adiabatic scalar inhomogeneities with high enough sound speed \cite{Barrow:2003ni}. Furthermore, the ES Universe was recently shown to be unstable to Bianchi type-IX spatially homogeneous perturbations in the presence of nontilted and tilted perfect fluids with $\rho + 3P>0$ \cite{arXiv:1108.3962} and for several kinds of matter fields sources (see \cite{Barrow:2009sj} and references therein).

The renewed interest in the ES Universe, besides its historical importance, comes from the Emergent Universe scenario \cite{Ellis:2002we}, an inflationary cosmological model in which it plays a crucial role as initial state. This model, in turn, suffers from a fine-tuning problem which is ameliorated when modifications to the cosmological equations of GR are present. For this reason, analogous solutions have been considered in several different modified gravity models \cite{Gergely:2001tn} and quantum gravity models \cite{Mulryne:2005ef,Parisi:2007kv,Boehmer:2009yz,Park:2009zra,Wu:2009ah,Canonico:2010fd}. Indeed, when dealing with modified cosmological equations, many new static solutions are present, whose stability properties, depending on the details of the single theory or family of theories taken into account, are substantially different from those of the classical ES solution of GR. In particular, neutrally stable solutions are present thus the fine-tuning problem is ameliorated but then a mechanism is needed to get out of the phase of infinite expansions and recollapses and to trigger the expanding phase of the Universe\cite{graceful}.

Here we study static cosmological solutions in the framework of a covariant Massive Gravity (MG) model recently proposed in \cite{derham10,derahm11}. In order to construct a consistent theory, nonlinear terms should be tuned to remove order by order the negative energy state in the spectrum \cite{boulware72}. The model under investigation follows from a procedure originally outlined in \cite{arkani03,creminelli05} and has been found not to show ghosts at the complete nonlinear level with an arbitrary reference metric \cite{Hassan:2011tf,Hassan:2011ea}.

The considered theory exploits several remarkable features. Indeed the graviton mass typically manifests itself on cosmological scales at late times thus providing a natural explanation of the presently observed accelerating phase \cite{car2012}. Moreover, the theory allows for exotic solutions in which the contribution of the graviton mass affects the dynamics at early times. Indeed, in contrast with GR where, in order to have static solutions, a cosmological constant term and a positive curvature term are needed in addition to a suitable perfect fluid source term, we find that in the considered MG theory it is possible to have static cosmological solutions only sourced by a perfect fluid. These solutions can be either unstable or neutrally stable and they exist even for spatially flat  (i.e. $\mathcal{K}=0$) cosmological models.

This paper is structured as follows. In Sec. \ref{massive}, the nonlinear MG framework considered in this work is shortly described and the modified Friedmann equations under investigation are introduced. In Sec. \ref{stab} static cosmological solutions are found, a linearized analysis is performed and the stability properties are discussed in details. In Sec. \ref{num} the dynamics near the fixed points is described using numerical integrations. The phase diagrams of the system are drawn both in the $(a, \dot{a})-$plane and $(H,\rho)-$plane. In Sec. \ref{concl}, some conclusions are eventually drawn. 
\section{Cosmological equations}\label{massive}

We consider the theory introduced in \cite{derham10,derahm11}. In the formalism afterwards used in \cite{massive}, the theory we refer to is defined on a four-dimensional pseudo-Riemannian manifold $(\mathcal{M},g)$ and the dynamics is determined by the action
\begin{equation}\label{totaction}
S=-\frac{1}{8\pi G}\int \left(\frac{1}{2} R+m^2  \mathcal{U}\right) d^{4}x \ + \ S_{m}, \nonumber
\end{equation}
where $G$ is the Newton gravitational constant and $R$ is the Ricci scalar while $S_m$ describes ordinary matter. The potential term, coupled through a mass term $m$, is defined by
\begin{eqnarray}\label{potential}
\mathcal{U}&=&\frac{1}{2} (K^2-K^\nu_\mu K^\mu_\nu)+\frac{c_3}{3!}\epsilon_{\mu\nu\rho\sigma}\epsilon^{\alpha\beta\gamma\sigma}K_\alpha^\mu K_\beta^\nu K_\gamma^\rho\nonumber\\ \nonumber
&&+\frac{c_4}{4!}\epsilon_{\mu\nu\rho\sigma}\epsilon^{\alpha\beta\gamma\delta}K_\alpha^\mu K_\beta^\nu K_\gamma^\rho K_\delta^\sigma, \nonumber
\end{eqnarray}
where $\epsilon_{\mu\nu\rho\sigma}$ is the Levi-Civita tensor density, $c_3$ and $c_4$ are arbitrary dimensionless real constants and 
$$
K^\mu_\nu=\delta^\mu_\nu-\gamma^\mu_\nu,
$$
$\gamma^\mu_\nu$ being defined by the relation
$$
\gamma^\mu_\sigma\gamma^\sigma_\nu=g^{\mu\sigma} f_{\sigma \nu}.
$$
with $f_{\sigma_\nu}$ a symmetric tensor field. The quantity $m_g=\hbar m/c$ is called the graviton mass.

We consider a Robertson-Walker Universe with three-dimensional spatial curvature $\mathcal{K}=0,\pm1$, described by the line element
\begin{eqnarray*}
ds^2&=&g_{\mu\nu}dx^\mu dx^\nu\\
&=&dt^2-a(t)^2\left[\frac{dr^2}{1-\mathcal{K}r^2} +r^2(d\theta^2+\sin(\theta)^2d\phi^2)\right].
\end{eqnarray*}
The first Friedmann equation has been written in \cite{chamseddine11} for generic values of the dimensionless constants $c_3$ and $c_4$ with the constraints imposed by Bianchi identities; it reads
\begin{eqnarray*}\label{mgfriedmann}
3\frac{\dot{a}^2+\mathcal{K}a^2}{a^2}&=&m^2 \left(4c_3+c_4-6+3C\frac{3-3c_3-c_4}{a}\right. \nonumber\\
&&\left.+3C^2\frac{c_4+2c_3-1}{a^2}-C^3\frac{c_3+c_4}{a^3}\right) + 8\pi G \rho
\end{eqnarray*}
where $C$ is an integrating constant. Matter couples minimally to gravity thus its equation of motion is
\begin{equation}\label{fluid}
\dot{\rho}+3H(\rho+p)=0
\end{equation}
with $H=\dot{a}/a$. From now on we assume a constant equation of state parameter $w$ thus $p=w\rho$.\\
Moreover, in the subsequent analysis the parameter space is reduced to the subset $c_3=-c_4$ since, as found in \cite{koyama11}, this is the simplest choice that presents a successful Veinshtein effect in the weak field limit.

For later purposes it is useful to rewrite Eq.(\ref{mgfriedmann}) as follows:
\begin{eqnarray}\label{friedmann}
H^{2}=\frac{\kappa}{3}\rho - \frac{\mathcal{K}}{a^2}+ \frac{m^2}{3} \left( A_1+\frac{A_2}{a}+\frac{A_3}{a^2 }\right)
\end{eqnarray}
where $\kappa=8 \pi G$,  $a/C \rightarrow a$ and
\begin{eqnarray*}
A_1&=&-3c_4-6\\
A_2&=&3\left(3 +2c_4 \right)\\
A_3&=&-3\left(1 +c_4 \right).
\end{eqnarray*}
The second Friedmann equation reads
\begin{eqnarray}\label{Ray}
\dot{H}&=& -\frac{\kappa}{2}\rho(1+w) + \frac{\mathcal{K}}{a^2} - \frac{m^2}{6} \left( \frac{A_2}{a}+2\frac{A_3}{a^2 }\right).  
\end{eqnarray}
 Making use of the Friedmann constraint Eq.(\ref{friedmann}) one can recast Eq.(\ref{Ray}) as a second order nonlinear differential equation in $a$ and its first and second derivatives
\begin{eqnarray*}\label{Ray2}
\frac{\ddot{a}}{a}&=&  \frac{m^2}{2}   \left[ (1+w)A_1 + \frac{2+3w}{3} \left(\frac{A_2}{a}\right)+\frac{1+3w}{3}\left(\frac{A_3}{a^2} \right) \right]  \nonumber \\
&& - \frac{1+3w}{2}\left( H^2+\frac{\mathcal{K}}{a^2} \right),
\end{eqnarray*}
which can be easily recast as a proper two-dimensional autonomous dynamical system by introducing the variables:
\begin{equation*}
q=a \qquad p=\dot{a}.
\end{equation*}
Thus, the system to be considered is the following:
\begin{eqnarray}
\dot{q}&=&p  \label{newsys1} \\
\dot{p}&=&\frac{m^2}{2} \left[  A_1 (1+w) q+A_2 \frac{2+3 w}{3}\right]+ \nonumber \\
&&\frac{-1}{2 q}(1+3 w) \left[ \frac{A_3}{3} m^2 + (\mathcal{K} +  p^2) \right]. \label{newsys2}
\end{eqnarray} 
The dynamics described by the above equations is globally Hamiltonian with respect to the symplectic structure
\begin{equation*}
\omega = q^{1+3w} dq \wedge dp,
\end{equation*}
which is singular in $q=0$.
Indeed
\begin{equation*}
i_X \omega = -d\mathcal{H}
\end{equation*}
where
\begin{equation} \label{Hamilton}
\mathcal{H}= \frac{q^{3(1+w)}}{2}\!\left[ \left(\frac{p}{q}\right)^2 \! + \! \frac{\mathcal{K}}{q^2} \!-\!\frac{m^2}{3}\left(A_1+\frac{A_2}{q}+\frac{A_3}{q^2}\right)\right],
\end{equation} 
and $i_X$ is the contraction operator with respect to the vector field
\begin{eqnarray*}
X&=&p\frac{\partial}{\partial q}+ \left(  \frac{m^2}{2} \left(  A_1 (w+1) q+A_2 \frac{2+3 w}{3}\right)+ \right.  \\ \nonumber
   && \left. \frac{-1}{2 q}(3 w+1) \left( \frac{A_3}{3} m^2 + (\mathcal{K} +  p^2) \right) \right)   \frac{\partial}{\partial p},
\end{eqnarray*}
which is singular in $q=0$ and $p=0$. The Hamilton's equations read
\begin{equation*}
\dot{q}=q^{-(1+3w)}\frac{\partial \mathcal{H}}{\partial p},  \qquad \dot{p}= -q^{-(1+3w)}\frac{\partial \mathcal{H}}{\partial q}.
\end{equation*} 

\section{Static solutions and their stability}\label{stab}
By imposing the condition $\dot{a}=\ddot{a}=\dot{\rho}=0$, Eq.(\ref{fluid}) is identically satisfied and the system of Eqs.(\ref{fluid})-(\ref{Ray}) reduces to an algebraic system in the unknowns $a$ and $\rho$. From this we get
\begin{eqnarray}
a_{\pm}&=&-\frac{m A_2 (2+3w) \pm \sqrt{\Omega} }{6 m (1+w)A_1} \label{static} \\
\rho_{\pm}&=&\frac{3}{\kappa} \left[\!\frac{\!\mathcal{K}}{a_{\pm}^2}-\frac{m^2}{3}\left(A_1+\frac{ A_2}{a_{\pm} }+\frac{A_3}{a_{\pm}^2}\! \right) \right] \label{rho}
\end{eqnarray}
with 
\begin{equation*}
\Omega\!=\!m^2 (2+3w)^2 A_2^2+\! 12 (3\mathcal{K}-\!m^2 A_3)(1+\!4w+\!3w^2)A_1.
\end{equation*}
Interesting enough, these solutions may exist not only for $\mathcal{K}=1$ as in GR, indeed the modified cosmological equations of MG  allow for static solutions also in both the $\mathcal{K}=0$ case and the $\mathcal{K}=-1$ case.

The stability analysis of the formerly presented solutions can be easily performed using standard dynamical system techniques.
It is easy to check that the solutions in Eqs.(\ref{static}) and (\ref{rho}) are stationary points of the dynamical system in Eqs.(\ref{newsys1}) and (\ref{newsys2}). Their stability is readily determined by looking at the eigenvalues of the Jacobian matrix evaluated at the stationary points.

Since the solutions we are interested in are static, once the eigenvalues of the linearized system are determined in general, we can impose $p=0$; they then reduce to a particularly simple form
\begin{equation}\label{eigenv}
\lambda_{1,2}=\pm \frac{\sqrt{\Sigma}}{\sqrt{2} q^2}
\end{equation}
with
\begin{eqnarray}\label{arg}
\Sigma&=&q^2\left[m^2 \left(-\frac{A_3}{3} (1+3w)+A_1 (1+w) q^2\right) \right. \nonumber \\
&&\left. +\mathcal{K}(1+3w) \right].
\end{eqnarray}
According to the sign of $\Sigma$ evaluated at the stationary points, one can have either a pair of real eigenvalues with opposite signs or a pair of purely imaginary eigenvalues with opposite signs. Thus, the fixed points are either neutrally stable \footnote{Lyapunov stable but not attracting \cite{Stro}} (i.e. center) when $\Sigma<0$ or unstable (i.e. a saddle) when $\Sigma>0$.

For the sake of simplicity we will first consider in details the simplest but interesting case $\mathcal{K}=0$ which shares all the relevant features of the other more involved ones.
\subsection{Case 1: $\mathcal{K}=0$}\label{K0}
In the case of spatially flat models the solutions in Eq.(\ref{static}) reduce to the simple expression
\begin{equation}\label{staticflat}
a_{1,2}=\frac{(2c_4+3)(2+3w) \pm \sqrt{(2c_{4}+3)^2 + (3w+2)^2-1}}{6(c_{4}+2)(1+w)},
\end{equation}
in which the explicit dependence on $c_4$ has been restored. The corresponding expression for the energy density in terms of the model parameters can be obtained replacing these solutions in Eq.(\ref{rho}). The conditions on the parameters for these solutions to exist can be found by simultaneously imposing $a_i>0$ and $\rho_i>0$. The results are reported in Table \ref{tab1}
\begin{table}[h!]
\begin{center}
\begin{tabular}{|c|c|c|}
\hline
Sol.&$w$& $c_4$ \\  \hline\hline
&$w<-1$&$c_4\neq-2,-1$ \\\cline{2-3}
  Sol.1&$ -1<w\leq-\frac{2}{3}$ & $c_4\neq -2 \cap c_4<-\frac{3+\sqrt{3(1+4w+3w^2)}}{2}$ \\ \cline{2-3}
  &$-\frac{2}{3}<w\leq-\frac{1}{3}$&$c_4<-2$\\\cline{2-3}
  &$w>-\frac{1}{3}$&$c_4<-2\quad c_4>-1$\\\hline
  Sol.2&$ -1<w<-\frac{2}{3}$ & $-2<c_4<-\frac{\sqrt{3(-1-4w-3w^2)}+3}{2}$ \\\hline
\end{tabular}
\caption{Existence conditions for the static solutions of the spatially flat model.}
\label{tab1}
\end{center}
\end{table}

The stability of the solutions is determined by evaluating the eigenvalues in Eq.(\ref{eigenv}) at these fixed points. As already observed, the problem simply consists in evaluating the sign of the function $\Sigma$ in Eq.(\ref{arg}). We find that the first solution is always unstable of the saddle type while the second is always (neutrally) stable, i.e., it is a center. In the latter case the linearized analysis is not sufficient to ensure that the solution is actually stable, indeed the hypotheses of the Hartman-Grobman theorem are not fulfilled being the point nonhyperbolic. 

To further analyze the second solution we use the Lyapunov's second method. Let us consider the function $\mathcal{H}$ as in Eq.(\ref{Hamilton}) with $\mathcal{K}=0$ and define
\begin{equation}
V(q,p)=\mathcal{H}(q,p)-\mathcal{H}(a_2,0).
\end{equation}
$V(q,p)$ is positive-definite in a neighborhood $U$ around the second solution, i.e.
\begin{eqnarray}\label{Lya}
V(q,p)&=&0 \quad for \quad \left(q=a_2,p=0\right)  \\
V(q,p)&>&0 \quad \forall\left(q,p\right) \in U \setminus \left(a_2,0\right),
\end{eqnarray} 
thus it is a good Lyapunov function candidate. Its time derivative is zero in a neighborhood of the fixed point i.e.
\begin{equation}\label{derLya}
\dot{V}(q,p) = \dot{\mathcal{H}} (q,p)= 0 \quad  \forall\left(q,p\right) \in U,
\end{equation} 
thus the second solution is proven to be stable. The result is also confirmed by numerically integrating the fully nonlinear system (see Sec.\ref{num}).

\begin{figure}[h!]
\centering
\subfigure[]{\label{flat1} \includegraphics[scale=0.38]{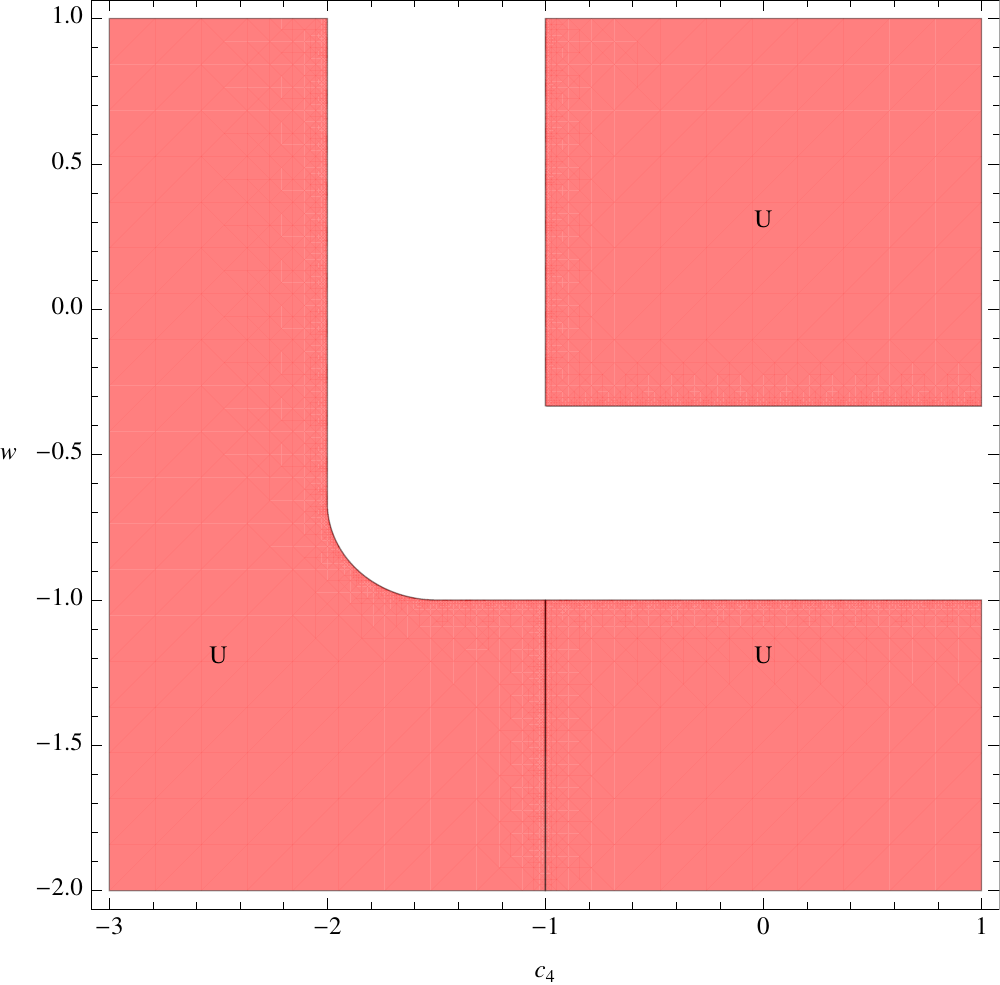}}
\subfigure[]{\label{flat2} \includegraphics[scale=0.38]{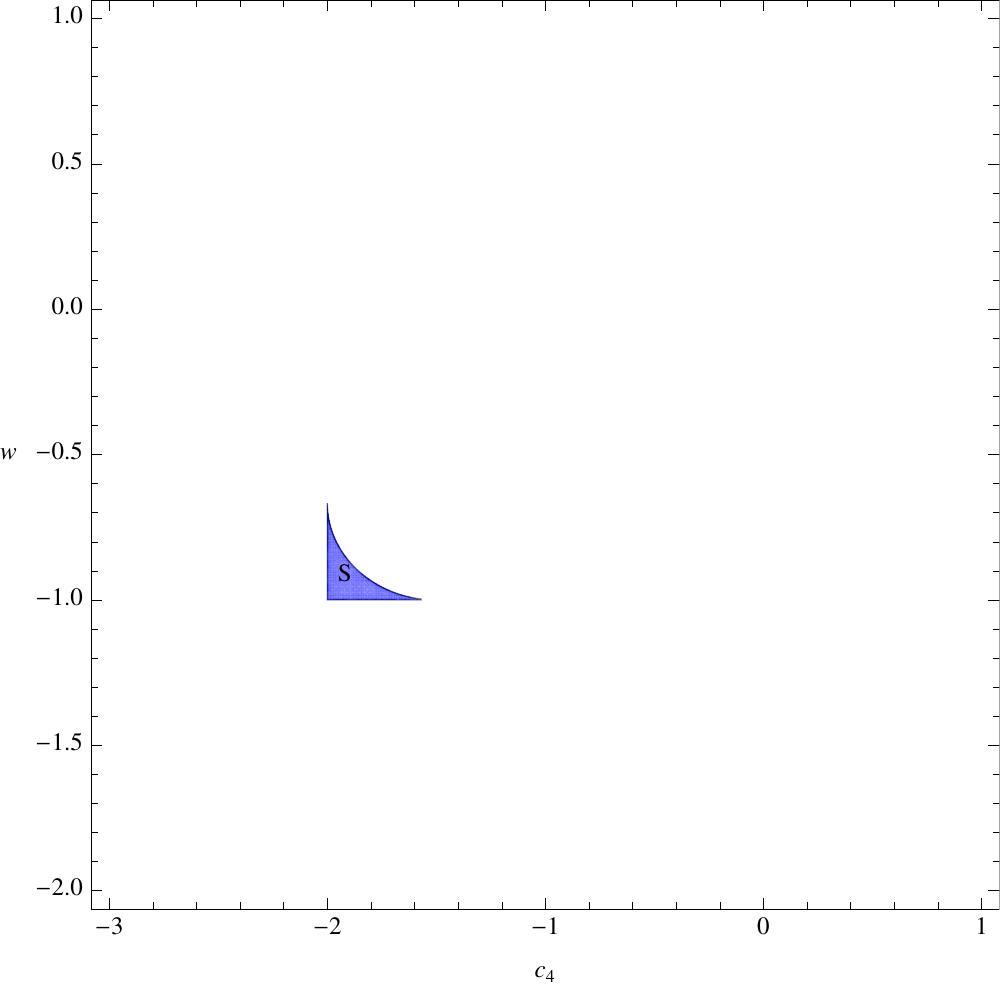}}
\caption{Stability conditions for the two static solutions of the $\mathcal{K}=0$ case in terms of the parameters $c_4$ and $w$. The stability properties of the two solutions in the considered region, $-3<c_4<1$ and $-2<w<1$, are always different. The first solution (a) is unstable (U, red); the second solution (b) is stable (S, blue).}
\label{stab12_flat}
\end{figure}

\begin{figure}[h!]
\begin{center}
\includegraphics*[scale=.80]{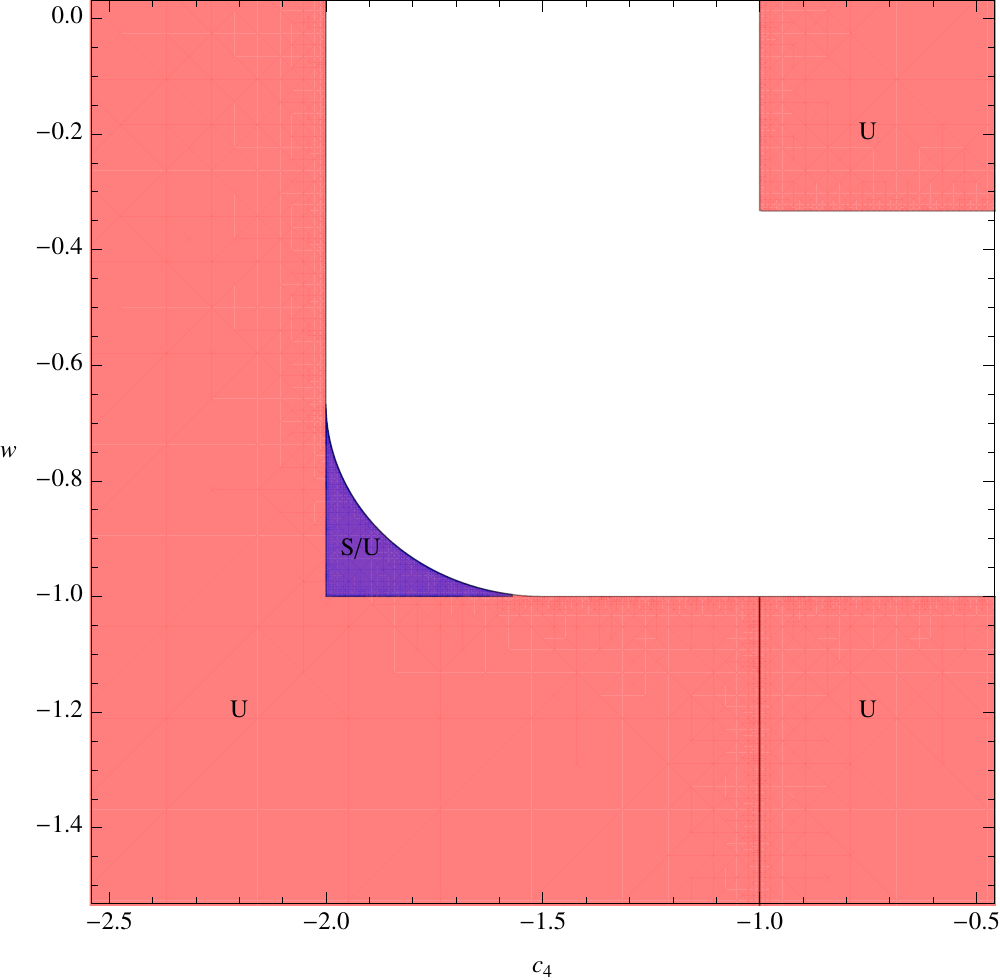}
\caption{{\small This figure is obtained by overlapping Figs.~\ref{flat1} and \ref{flat2}; it shows a region (U/S, purple color) of the parameters space allowing both solutions to exist, the first being unstable while the second being stable. }} \label{flat}
\end{center}
\end{figure}

Figs.~\ref{flat1} and \ref{flat2} show regions in the $(c_4,w)-$plane in which Sol.1 and Sol.2 are admitted, respectively; the former being unstable while the latter being stable. Such regions have a nonvanishing intersection. More precisely, as shown in Fig.~\ref{flat}, the region in Fig.~\ref{flat2} is completely contained by the region in Fig.~\ref{flat1}. For $c_4$ and $w$ in this region, both static solutions are admitted; otherwise either the unstable solution only or no static solutions are admitted. 

It is interesting to note that the phase space of the system exploits relevant changes in its qualitative structure according to the parameters values. In particular, the system undergoes bifurcations that can be singled out, for instance, by looking at the eigenvalues characterizing  the stability of the fixed points.

\begin{figure}[h!]
\centering
\subfigure[]{\label{flat1_bif} \includegraphics[scale=0.38]{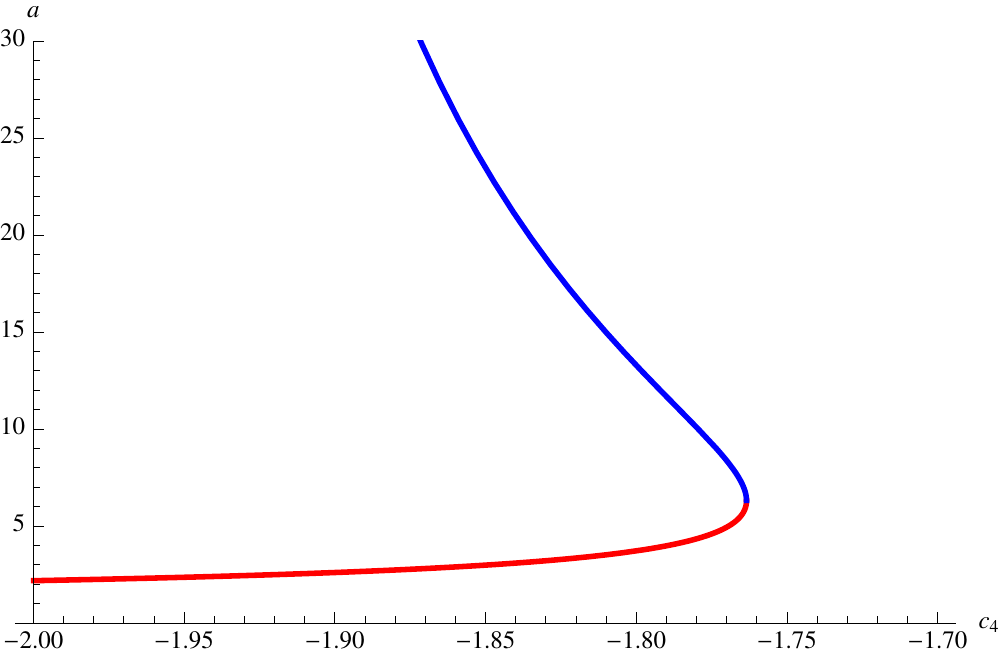}}
\subfigure[]{\label{flat2_bif} \includegraphics[scale=0.38]{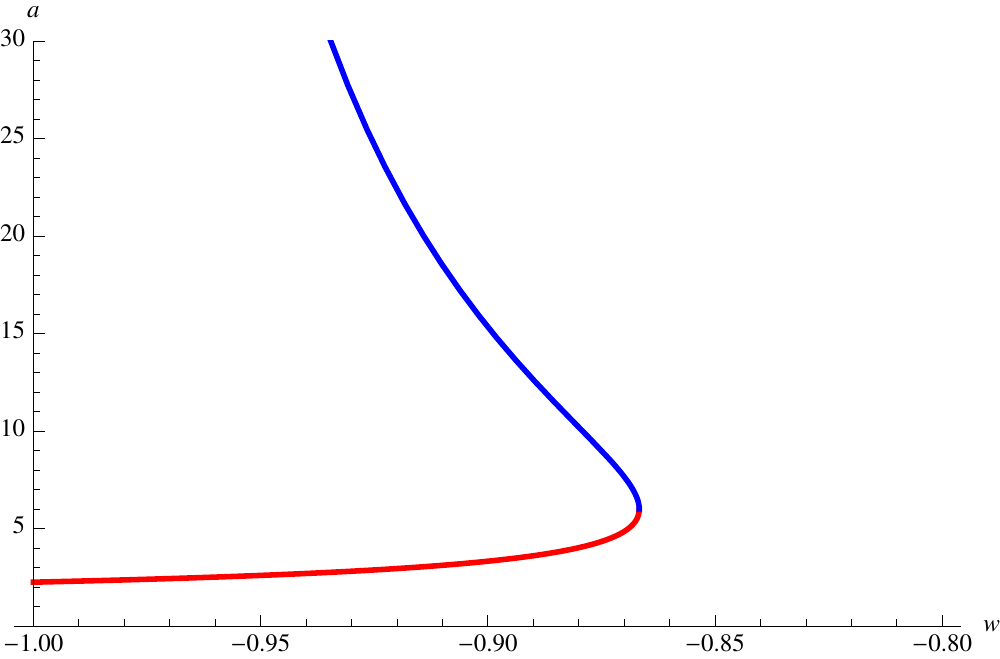}}
\caption{{\small Scale factor value of the static solutions ($a_{1,2}$) as a function of (a) $c_4$ and (b) $w$. The parameters have been arbitrarily chosen (i.e. $\kappa =1$, $m=1$, $w=0.08$, $c_4=0.6$) in order to show a typical bifurcation. The upper (blue) curve and the lower (red) curve correspond to the stable and unstable solution respectively.}} \label{flat_bif}
\end{figure}

In Fig.~\ref{flat_bif} an example of bifurcation is depicted. Fig.~\ref{flat1_bif} is obtained by varying the parameter $c_4$ while keeping the parameter $w$ fixed. The value of the scale factor corresponding to each static solution varies until the fixed points annihilate exploiting a saddle-center bifurcation. Fig.~\ref{flat2_bif} shows a similar behavior; it is obtained by varying the parameter $w$ while keeping the parameter $c_4$ fixed.
Further details about this feature are discussed in the following sections (see Sec.\ref{num}).
\subsection{Case 2: $\mathcal{K}=1$}\label{K1}
Spatially closed models can be analyzed following the formerly described procedure, that is imposing the positivity of $a$ and $\rho$ and then studying the sign of the function $\Sigma$ in Eq.(\ref{arg}). The analytical expression of the resulting ranges in terms of the parameters $w$ and $c_4$ is quite cumbersome and not particularly illuminating, for this reason we do not report it explicitly. We just remark that, as in the spatially flat case (Sec. \ref{K0}), at most two static solutions are admitted, one being unstable, the other being neutrally stable, in contrast with GR where only one unstable solution of the saddle type is admitted. 

\begin{figure}[h!]
\centering
\includegraphics*[scale=.80]{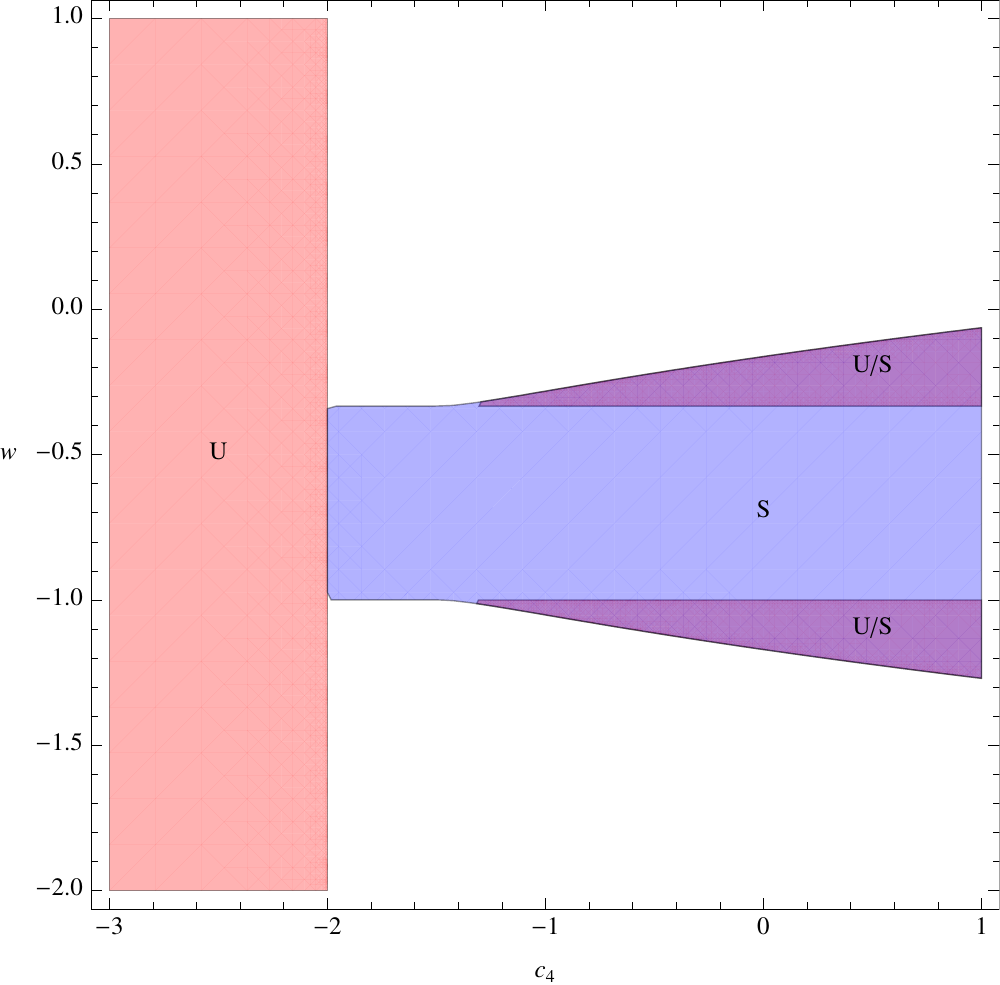}
\caption{{\small Stability conditions for the $\mathcal{K}=1$ case in terms of the parameters $c_4$ and $w$. The regions of existence of the two solutions partially overlap but the character of their stability is different, the first solution being unstable (U, red), the second solution being stable (S, blue). For parameters in the two triangular regions (U/S, in purple color) both solutions can exist.}} \label{closed}
\end{figure}

In the considered range, $-3<c_4<1$ and $-2<w<1$, the existence region of the first solution consists of three parts while the existence region of the second solution is compact (see Fig.~\ref{closed}). These regions have two intersections, which means that for values of the parameter within such regions, both solutions are admitted simultaneously, otherwise either only one of the two solutions is admitted or no solutions are admitted. This also indicates that in the spatially closed models the phase space has a richer structure and several bifurcations can occur. For instance, bifurcations diagrams similar to those found in the spatially flat case  (depicted in Fig.~\ref{flat_bif}) can be drawn.
\subsection{Case 3: $\mathcal{K}=-1$}
As in the case of spatially closed models, the analytical expression defining the ranges of existence of the two solutions for spatially open models is quite cumbersome. As in both the previously analyzed cases, the system in Eqs.~(\ref{newsys1}) and (\ref{newsys2}) admits two static solutions, one being unstable, the other being neutrally stable, in contrast with GR where only one static solutions is present\footnote{A static solution with open spatial sections is admitted in GR, it is unstable of the saddle type and requires very peculiar conditions, namely, a negative cosmological constant and a phantom fluid \cite{Canonico:2010fd}}. 

\begin{figure}[h!]
\centering
\includegraphics*[scale=.80]{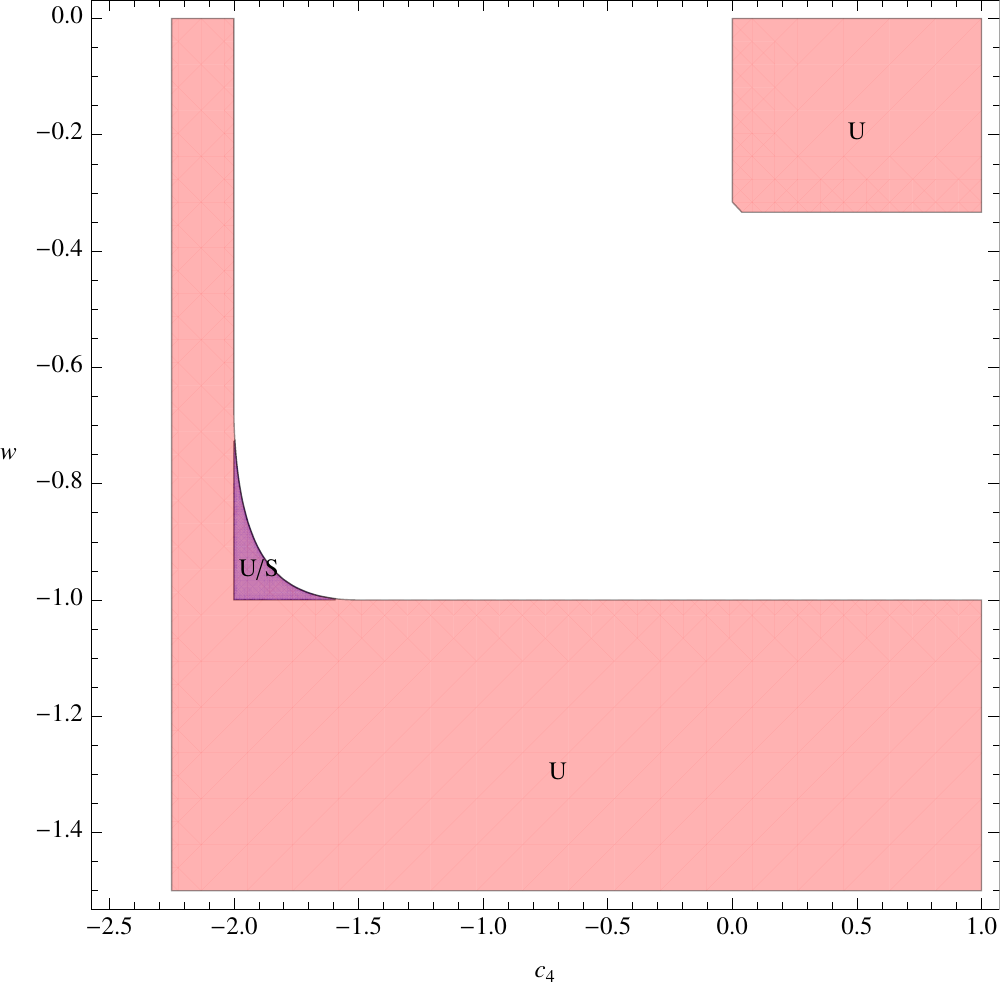}
\caption{{\small Stability conditions for the $\mathcal{K}=-1$ case in terms of the parameters $c_4$ and $w$. The regions of existence of the two solutions partially overlap but the character of their stability is always different. The first solution is unstable (U, red) and the second solution is stable (S, blue). There is a region of the parameter space (U/S, purple color) for which both solutions can exist.}} \label{open}
\end{figure}

In the considered range, $-3<c_4<2$ and $-2<w<1$, the existence region of the first solution consists of two parts while the existence region of the second solution is compact  (see Fig.~\ref{open}). The existence regions have one intersection; for $c_4$ and $w$ in this region, both static solutions are admitted, otherwise either the unstable solution only is admitted or no static solution is admitted. The result is very similar to that obtained for open models but now the system does not admit solutions for $c_4$ smaller than $\approx -2.25$. This result also suggests that, crossing the boundary of the region, where both solutions are admitted, bifurcations can occur. For instance, bifurcations diagrams similar to those found in the spatially flat case  (depicted in Fig.~\ref{flat_bif}) can be easily drawn.
\section{Numerical integration}\label{num}
In this section we discuss the properties of the considered solutions and the phase space of MG by performing numerical integrations of the system in Eqs.~(\ref{newsys1}) and (\ref{newsys2}). This procedure allows us to check the results of the linearized stability analysis for the nonhyperbolic fixed points and provides further interesting physical information. For the seek of simplicity, we just consider the spatially flat case.

\begin{figure}[h!]
\centering
\includegraphics*[scale=.80]{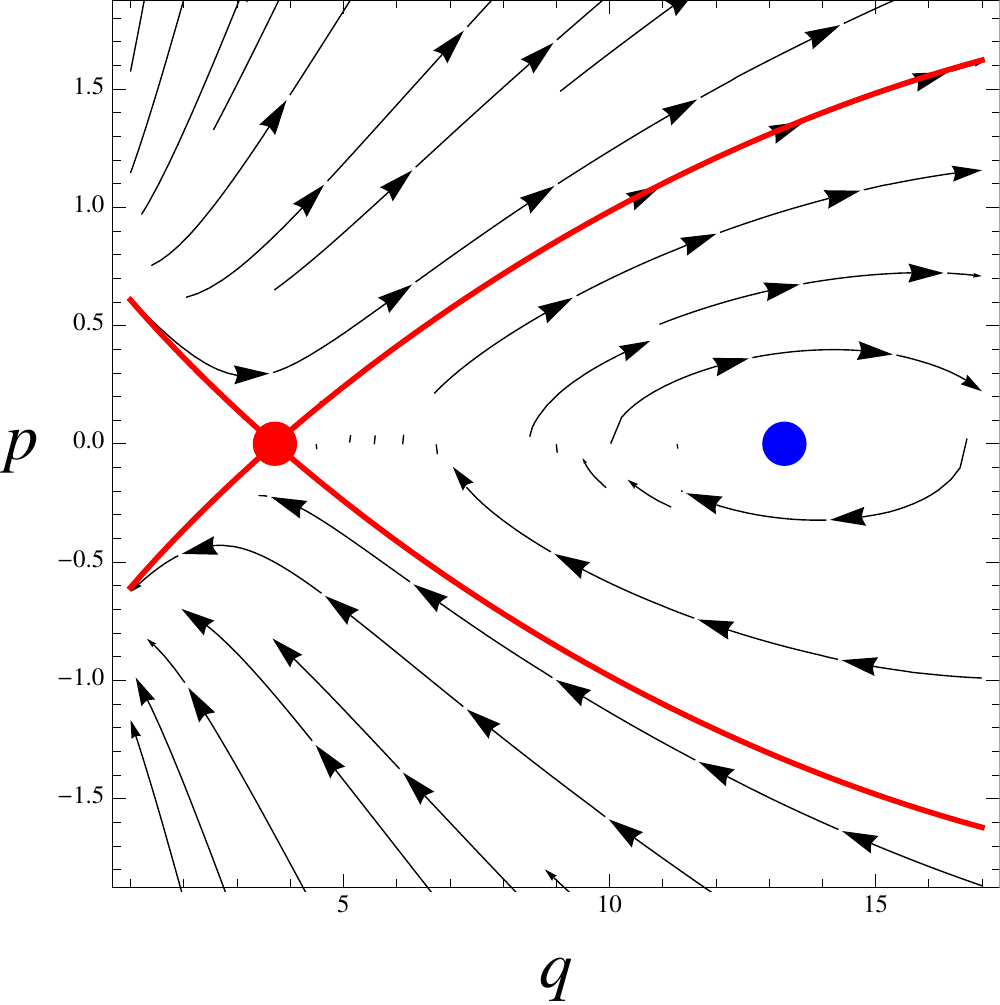}
\caption{{\small Phase space of the dynamical system for the $\mathcal{K}=0$ case. This figure is obtained by integrating the nonlinear system for initial conditions close to the fixed points. The (red) point on the left represents the unstable static solution while the (blue) point on the right represents the stable static solution. The continuous (red) curves intersecting the unstable point are parts of the separartix.}} \label{x1x2}
\end{figure}

Fig.~\ref{x1x2} shows a region of the $(q, p)-$plane for the system in Eqs.(\ref{newsys1}) and (\ref{newsys2}). The (red) point on the left represents the unstable fixed point, the (blue) point on the right represents the nonhyperbolic fixed point. Arrows represent the orbits obtained evolving initial conditions. The (red) continuous curve is a separatrix, it marks the boundary of regions where the dynamical behavior of the system is different. Initial conditions, belonging to the region enclosed by the separatrix, evolve producing closed orbits which remain close to the nonhyperbolic fixed point confirming that it is neutrally stable.

More physical insights can be gained by rewriting the systems in terms of different variables, namely $H$ and $\rho$. To this aim one first has to consider the original systems of Eqs.~(\ref{fluid})-(\ref{Ray}). The Friedmann constraint in Eq.~(\ref{friedmann}) can be locally solved in order to express the scale factor $a$ in terms of $H$ and $\rho$, then, by substituting in Eq.(\ref{Ray}), one gets a new equation for $\dot{H}$ which, together with Eq.~(\ref{fluid}), is a two-dimensional autonomous dynamical system.

\begin{figure}[h!]
\centering
\subfigure[]{\label{bif_rho_H1} \includegraphics[scale=0.38]{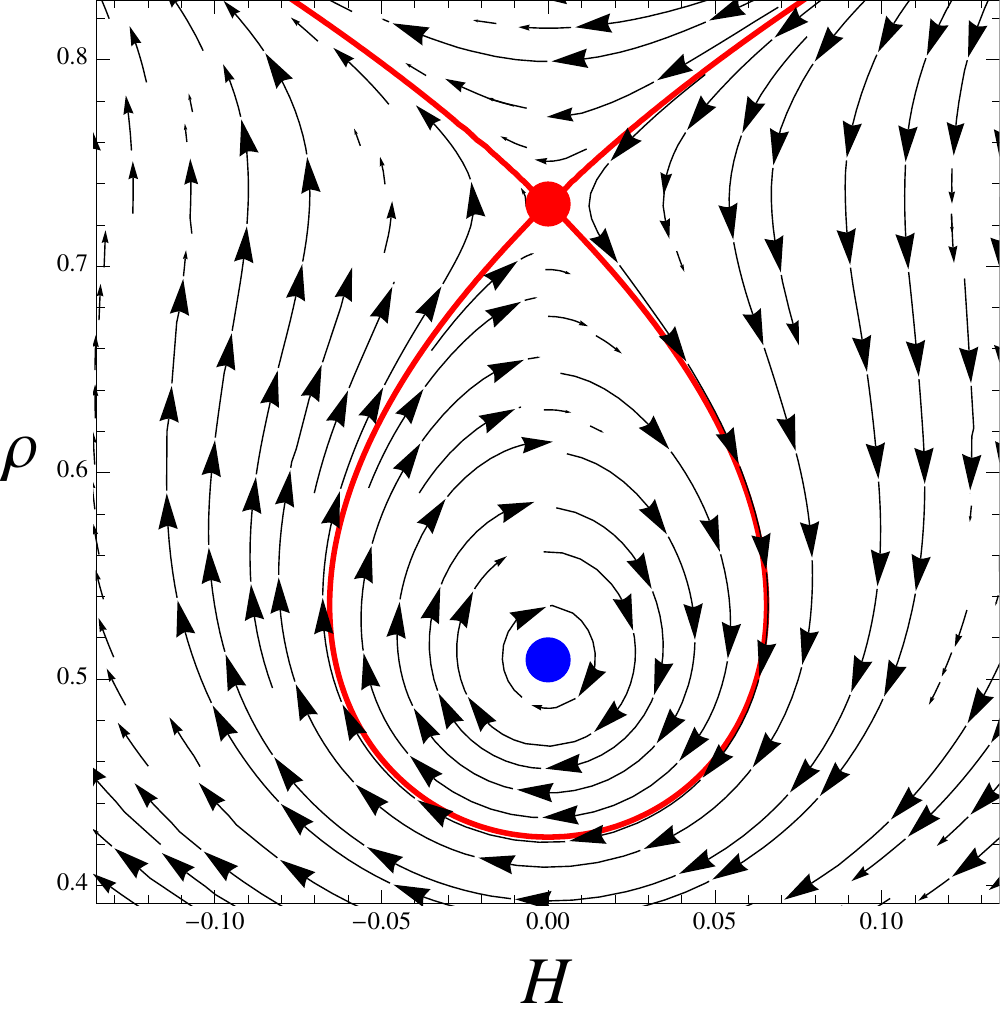}}
\subfigure[]{\label{bif_rho_H2} \includegraphics[scale=0.38]{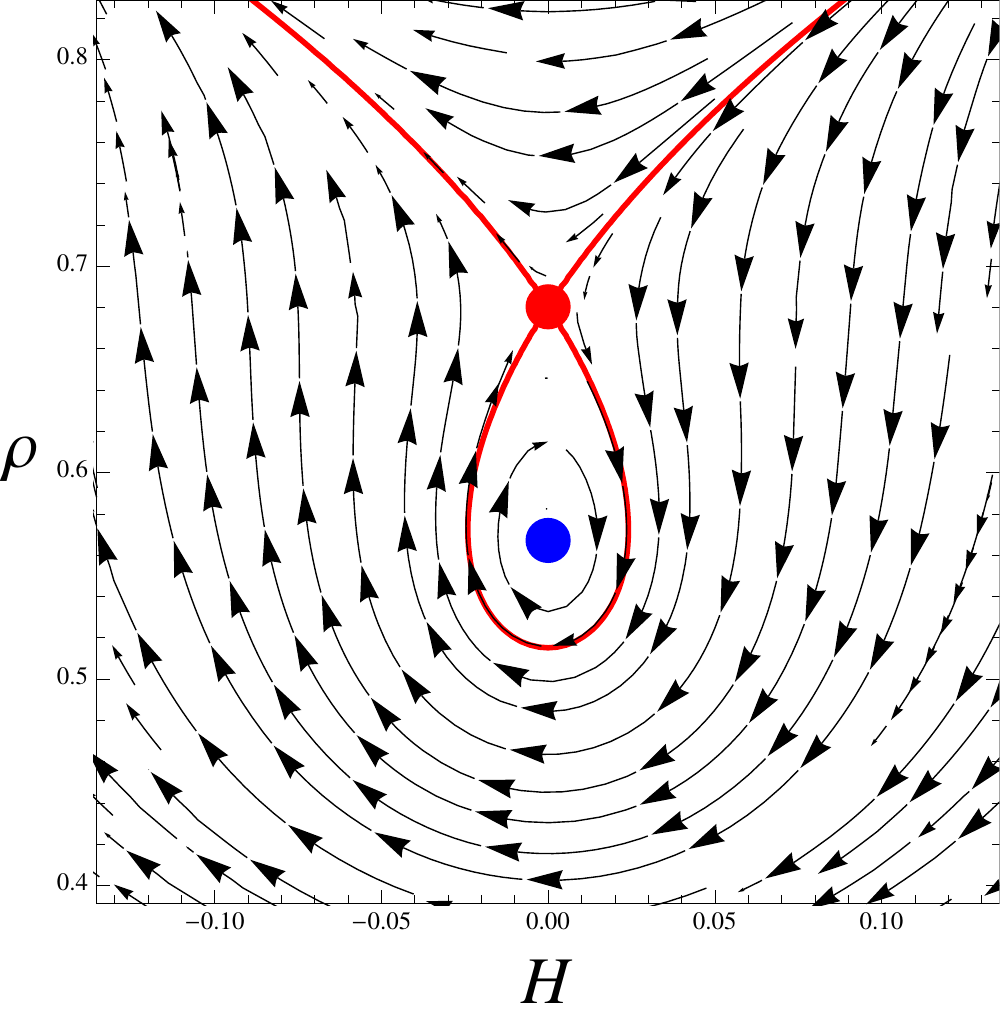}}\\
\subfigure[]{\label{bif_rho_H3} \includegraphics[scale=0.38]{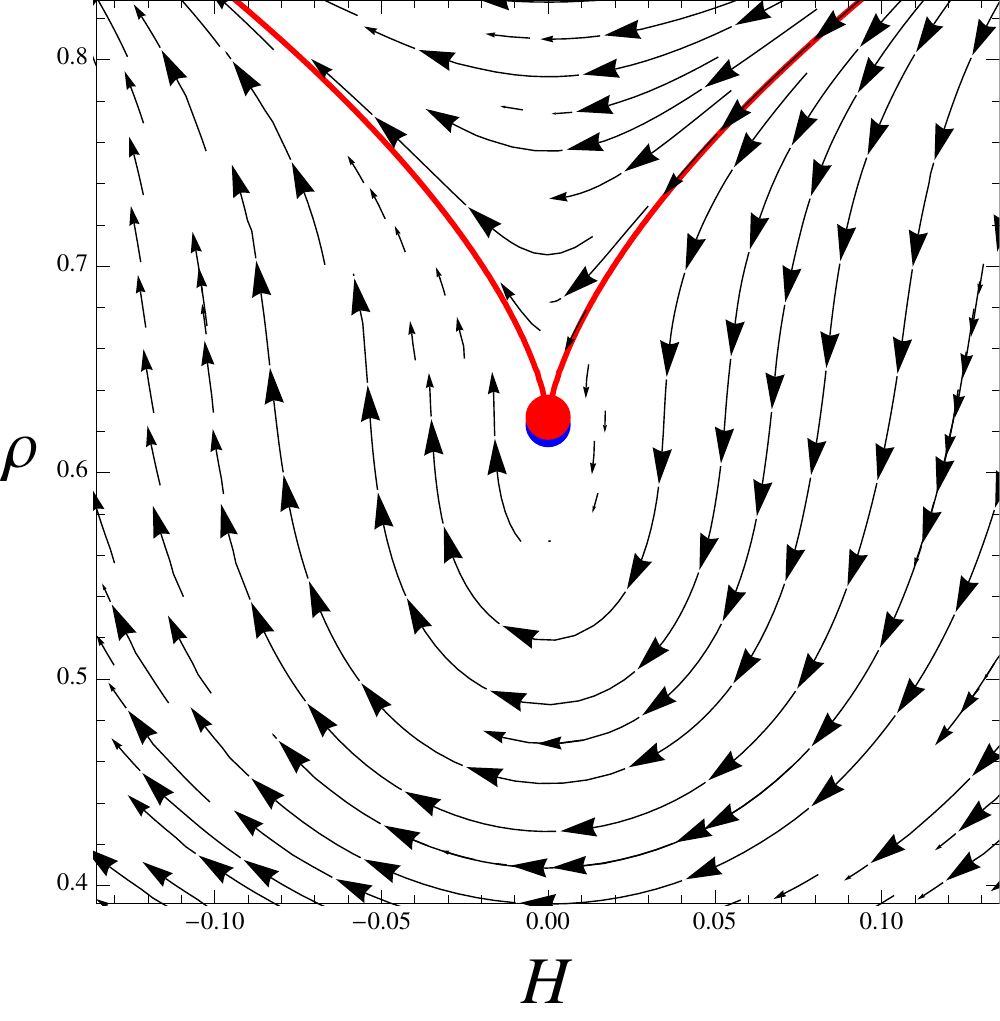}}
\subfigure[]{\label{bif_rho_H4} \includegraphics[scale=0.38]{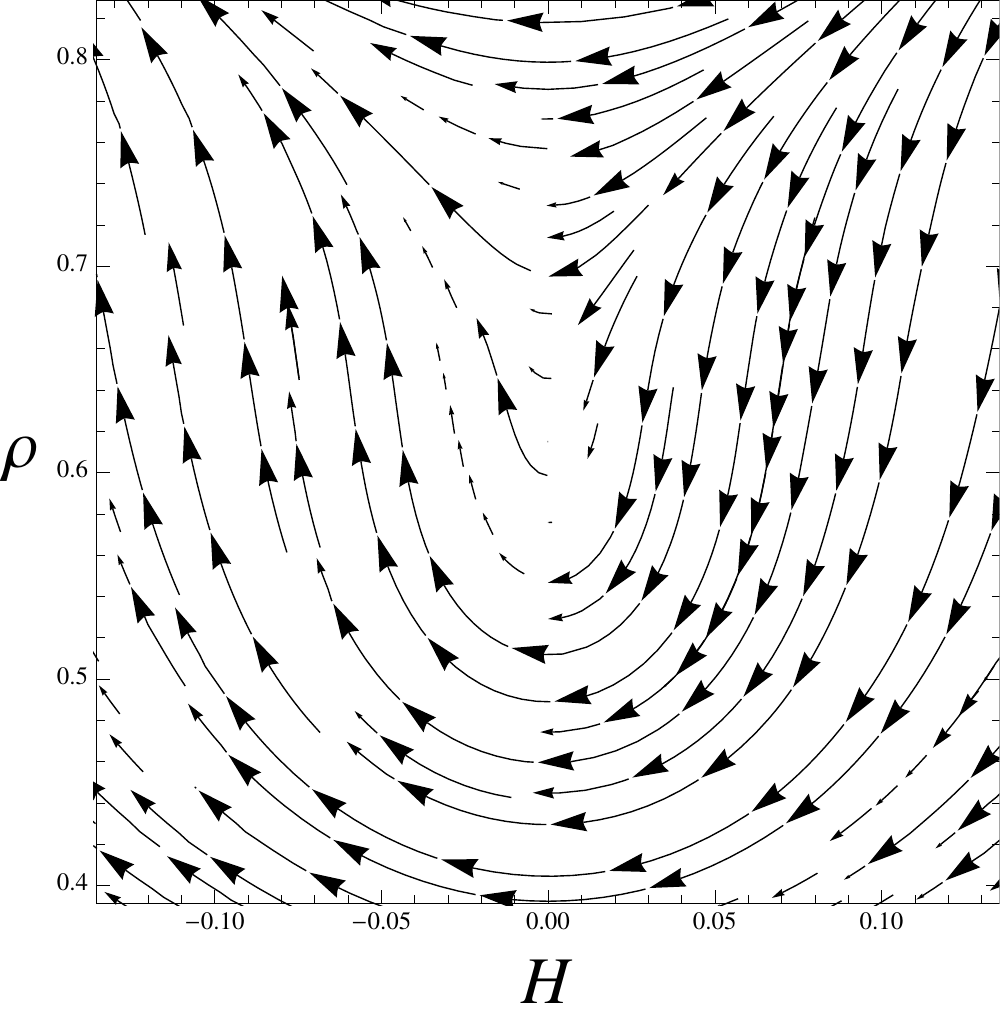}}
\caption{ {\small Dynamical behaviour of the system in the $(H,\rho)-$plane for initial conditions close to the static solutions. The continuous (red) line is a separatrix, a curve which crosses the unstable (upper, red) fixed point, enclosing the stable region around the neutrally stable (lower, blue) fixed point. The phase space portraits are drawn using always the same arbitrary values for the parameter and varying the the equation of state parameter, i.e. (a) $w=-0.88$, (b) $w=-0.87$, (c) $w=-0.86667$, (d) $w=-0.865$.}}
\label{bif_rho_H}
\end{figure}

Fig.~\ref{bif_rho_H} shows the phase space portrait for different suitably chosen values of the model parameters. The (red) continuous line is a separatrix enclosing the stable region containing the static solution. The closed orbits represent cosmological models characterized by an infinite sequence of bouncing and recollapsing epochs. The pictures Figs.~\ref{bif_rho_H1}~-~\ref{bif_rho_H4} are obtained by varying the value of the equation of state parameter $w$ and keeping constant the other model parameter. $w$ plays the role of bifurcation parameter indeed, according to its values, the distance between the fixed points decreases until the two solutions collapse on the same point and disappear exploiting a saddle-center bifurcation.
\section{Conclusions}\label{concl}
We have considered static solutions in the cosmological sector \cite{chamseddine11} of a recently proposed theory of massive gravity \cite{derham10}. We have shown that the effect of a massive graviton is to enrich the phase space of the cosmological equations, enlarging the ranges of existence of static solutions and affecting their stability properties. The solutions found exhibit stability properties rather different from those of the standard ES solution of GR, which requires a positive cosmological constant and a positive spatial curvature in addition to a perfect fluid with an equation of state parameter $w>-1/3$.

Two kinds of solutions are present: neutrally stable solutions and unstable solutions (of the saddle type). Summing up, for spatially closed ($\mathcal{K}=1$) models three cases are possible: $i)$ both the unstable and the neutrally stable solutions are admitted, $ii)$ either the unstable solution or the neutrally stable solution is admitted, or $iii)$ no static solutions are admitted. For spatially flat ($\mathcal{K}=0$) and open ($\mathcal{K}=-1$) models, three cases are possible: $i)$ both the unstable and the neutrally stable solutions are admitted, $ii)$ only the unstable solution is admitted, or $iii)$ no static solutions are admitted. Notice that, in the considered region of the parameter space, the neutrally stable solution requires a negative equation of state parameter $w$; in particular it must be $-1<w<-2/3$ in the $\mathcal{K}=0,-1$ cases and $-1.2<w<0$ in the $\mathcal{K}= 1$ case.

Our result implies the possibility of constructing models in which the Universe oscillates indefinitely about an initial static state, thus the fine-tuning problem suffered by the emergent Universe scenario in GR \cite{Ellis:2002we} is ameliorated when MG modifications are taken into account. On the other hand, this result raises the question of finding a mechanism in order to break the regime of infinite oscillations able to enter the current expanding phase undergone by the Universe \cite{graceful}. This result can be achieved by varying one of the model parameters, namely the equation of state parameter $w$ and the dimensionless parameter $c_4$ due to MG modification, in order for the system to undergo a bifurcation thus chancing the topological structure of the phase space. Such a mechanism has no dynamical explanation thus it looks quite unsatisfactory as well. Moreover, a full-fledged stability analysis against nonhomogeneous and/or nonisotropic modes would probably reveal other instabilities as it happens in GR but such an analysis is well beyond the scope of this paper.

\begin{acknowledgements}
L.P. would like to thank R. Maartens for thoughtful advice and continuous encouragement. This work is partially supported by the Italian Ministero Istruzione Universit\`a e Ricerca (MIUR) through the PRIN 2008 grant and by the INFN/MICINN collaboration, Grant No.AIC10-D-000581.  L.P. is partially funded by Agenzia Spaziale Italiana (ASI).
\end{acknowledgements}


\begin{thebibliography}{99}

\bibitem{Hawking:1973uf}
  S.~W.~Hawking and G.~F.~R.~Ellis,
  ``The Large scale structure of space-time,''
{\it  Cambridge University Press, Cambridge, 1973}

\bibitem{Edd:1930}
A. S. Eddington, Mon.\ Not.\ Roy.\ Astron.\ Soc.\  {\bf 90}, 668 (1930).

\bibitem{Barrow:2003ni}
  J.~D.~Barrow, G.~F.~R.~Ellis, R.~Maartens and C.~G.~Tsagas,
  Class.\ Quant.\ Grav.\  {\bf 20} (2003) L155

\bibitem{Barrow:2009sj}
  J.~D.~Barrow and C.~G.~Tsagas,
  Class.\ Quant.\ Grav.\  {\bf 26} (2009) 195003.

\bibitem{arXiv:1108.3962}  
  J.~D.~Barrow and K.~Yamamoto,
  Phys.\ Rev.\  D {\bf 85}  (2012) 083505.
 
\bibitem{Ellis:2002we}
  G.~F.~R. Ellis  and R. Maartens,
  Class. Quant. Grav.  {\bf 21} (2004) 223-232;
  G.~F.~R. Ellis, J. Murugan  and C.~G. Tsagas,
  Class. Quant. Grav.  {\bf 21} (2004) 233-249;
  S. Mukherjee, B.~C. Paul, N.~K. Dadhich, S.~D. Maharaj and A. Beesham,
  Class. Quant. Grav.  {\bf 23} (2006) 6927-6933.
  
\bibitem{graceful}
 J.~E.~Lidsey and D.~J.~Mulryne,
  Phys.\ Rev.\  D {\bf 73} (2006) 083508;
  J.~E.~Lidsey, D.~J.~Mulryne, N.~J.~Nunes and R.~Tavakol,
  Phys.\ Rev.\  D {\bf 70} (2004) 063521.
  
\bibitem{Gergely:2001tn}
  L.~A.~Gergely and R.~Maartens,
  Class.\ Quant.\ Grav.\  {\bf 19}, 213 (2002);
  A. Gruppuso, E. Roessl and M. Shaposhnikov, JHEP {\bf 011} (2004) 0408;
  S. S. Seahra, C. Clarkson and R. Maartens, Class. Quant. Grav. {\bf 22} (2005) L91;
  C. Clarkson and S. S. Seahra, Class. Quant. Grav. {\bf 22} (2005) 3653;
  C. G. B\"{o}hmer, Class. Quant. Grav. {\bf 21} (2004) 1119; 
  T.~Clifton and J.~D.~Barrow,
  Phys.\ Rev.\  D {\bf 72}, 123003 (2005);  
  C.~G.~B\"{o}ehmer, L.~Hollenstein and F.~S.~N.~Lobo,
  Phys.\ Rev.\  D {\bf 76}, 084005 (2007);
  R. Goswami, N. Goheer and P. K. S. Dunsby, Phys. Rev. {\bf D78} (2008) 044011;
  N. Goheer, R. Goswami and P. K. S. Dunsby, Class. Quant. Grav. {\bf 26} (2009) 105003;
  S.~S.~Seahra and C.~G.~Boehmer,
  Phys.\ Rev.\  D {\bf 79} (2009) 064009;
  C. G. B\"{o}hmer and F. S. N. Lobo, Phys. Rev. {\bf D79} (2009) 067504;
  C.~G.~B\"{o}ehmer, L.~Hollenstein, F.~S.~N.~Lobo and S.~S.~Seahra,
  arXiv:1001.1266 [gr-qc];
  K.~Zhang, P.~Wu and H.~W.~Yu,
  Phys.\ Lett.\  B {\bf 690} (2010) 229;
  P.~Wu and H.~Yu,
  Phys.\ Lett.\  B {\bf 703}, 223 (2011).
  
\bibitem{Mulryne:2005ef}
  D.~J.~Mulryne, R.~Tavakol, J.~E.~Lidsey and G.~F.~R.~Ellis,
  Phys.\ Rev.\  D {\bf 71}, 123512 (2005).

 \bibitem{Parisi:2007kv}
  L.~Parisi, M.~Bruni, R.~Maartens and K.~Vandersloot,
  Class.\ Quant.\ Grav.\  {\bf 24} (2007) 6243.
  
\bibitem{Park:2009zra}
  M.~i.~Park,
  JHEP {\bf 0909} (2009) 123.
  
\bibitem{Boehmer:2009yz}
  C.~G.~Boehmer and F.~S.~N.~Lobo,
  Eur.\ Phys.\ J.\  C {\bf 70}, 1111 (2010).

\bibitem{Wu:2009ah}
  P.~Wu and H.~W.~Yu,
  Phys.\ Rev.\  D {\bf 81}, 103522 (2010).
  
\bibitem{Canonico:2010fd}
  R.~Canonico and L.~Parisi,
  Phys.\ Rev.\  D {\bf 82}, 064005 (2010).

\bibitem{derham10}
 C.~de Rham, G.~Gabadadze,
  Phys.\ Rev.\  {\bf D82 } (2010)  044020.
 
\bibitem{derahm11}
  C.~de Rham, G.~Gabadadze, A.~J.~Tolley,
  Phys.\ Rev.\ Lett.\  {\bf 106 } (2011)  231101.

\bibitem{boulware72}
 D.G. Boulware  and S. Deser,
 Phys. Rev. D {\sl 6}, 3368 (1972).

\bibitem{arkani03}
 N.~Arkani-Hamed, H.~Georgi, M.~D.~Schwartz,
  Annals Phys.\  {\bf 305 } (2003)  96-118.

\bibitem{creminelli05}
  P.~Creminelli, A.~Nicolis, M.~Papucci, E.~Trincherini,
  JHEP {\bf 0509 } (2005)  003.

\bibitem{Hassan:2011tf}
  S.~F.~Hassan, R.~A.~Rosen and A.~Schmidt-May,
  JHEP {\bf 1202} (2012) 026.

\bibitem{Hassan:2011ea}
  S.~F.~Hassan and R.~A.~Rosen,
  JHEP {\bf 1204} (2012) 123.
    
\bibitem{car2012}
  V.~F.~Cardone, N.~Radicella, L.~Parisi
  Phys.\ Rev.\  {\bf D85 }, 124005 (2012).
    
\bibitem{massive}
T.~M.~Nieuwenhuizen, Phys.\ Rev.\  {\bf D84 }, (2011) 024038.

\bibitem{chamseddine11}
 A.~H.~Chamseddine, M.~S.~Volkov,
  Phys.\ Lett.\  {\bf B704 } (2011)  652-654.

\bibitem{koyama11}
 K.~Koyama, G.~Niz, G.~Tasinato,
  Phys.\ Rev.\  {\bf D84 } (2011)  064033.
  
\bibitem{Stro}
S. H. Strogatz, {\it Nonlinear Dynamics And Chaos: With Applications To Physics, Biology, Chemistry, And Engineering}, Addison-Wesley, Reading, MA (1994).
\end{thebibliography}
\end{document}